\documentclass[10pt,epsfig,graphics,mathbbm]{article}

\usepackage{amsmath,amsfonts,amssymb,graphics,graphicx,epsfig,color,times,bbm}

\newcommand{\me}{\mathrm{e}}
\newcommand{\mi}{\mathrm{i}}

\newcommand{\cc}{\mathbbm{C}}
\newcommand{\nn}{\mathbbm{N}}
\newcommand{\rr}{\mathbbm{R}}
\newcommand{\id}{\mathbbm{1}}

\newtheorem{theorem}{Theorem}
\newtheorem{lemma}{Lemma}
\newtheorem{corollary}{Corollary}

\newtheorem{example}{Example}
\newtheorem{assumption}{Assumption}

\newcommand{\proof}{{\em Proof. }}
\newcommand{\qed}{\hfill$\square$\par\vskip24pt}

\begin{document}

\title{\Large\bf Correlations, spectral gap, and entanglement \\
in harmonic quantum systems on generic lattices}

\author{\large
M.\ Cramer$^{1}$ and J.\ Eisert$^{2,3}$}

\date{}
\maketitle

\vspace*{-.6cm}

\centerline{\it\footnotesize 
1 Institut f{\"u}r Physik, Universit{\"a}t Potsdam,
Am Neuen Palais 10, D-14469 Potsdam, Germany}

\centerline{\it\footnotesize  
2 QOLS, Blackett Laboratory, 
Imperial College London,
Prince Consort Road, London SW7 2BW, UK}

\centerline{\it\footnotesize  
3 Institute for Mathematical Sciences, Imperial College London,
Prince's Gardens, London SW7 2PE, UK}

\begin{abstract}
We investigate the relationship between the gap between
the energy of the ground state and the first excited
state and the decay of correlation
functions in harmonic lattice systems. 
We prove that in gapped systems, the 
exponential decay of correlations follows for both the 
ground state and thermal states. Considering the converse 
direction, we show
that an energy gap can follow from algebraic decay
and always does for exponential decay.
The underlying lattices are described as general
graphs of not necessarily integer dimension, including
translationally invariant instances of cubic lattices
as special cases. Any 
local quadratic couplings in position and momentum 
coordinates are allowed for, leading to quasi-free
(Gaussian) ground states. 
We make use of methods
of deriving bounds to 
matrix functions of banded matrices corresponding
to local interactions on general graphs. 
Finally, we give an explicit entanglement-area
relationship 
in terms of the energy gap for arbitrary, not necessarily 
contiguous regions on lattices characterized by general graphs.  
\end{abstract}

\section{Introduction}
On physical grounds, one expects that correlations in 
ground states of 
gapped many-body systems -- quantum systems for which
the energy of the lowest exitation is strictly larger than the ground
state energy -- decay 
exponentially. This implication of the energy gap to the 
exponential decay of two-point equal-time
correlation functions is a well-established observation on 
non-critical
quantum many-body systems 
\cite{Sachdev,LSM,Haldane,Hastings,HastingsFermion,Hastings2,Koma,Nachtergaele,Fredenhagen}.
Yet, surprisingly perhaps, rigorous proofs of this implication
for spin-models in higher dimensional cubic or general lattices 
have been found only very recently. Notably, Ref.\ 
\cite{Hastings2} reexamines this
question for finite-dimensional constituents, which has been
generalized to 
the case of general lattices defined by graphs 
that may have any non-integer dimension in Refs.\ \cite{Hastings,Koma}.

\begin{figure}
\begin{center}
\includegraphics[width=0.8\textwidth]{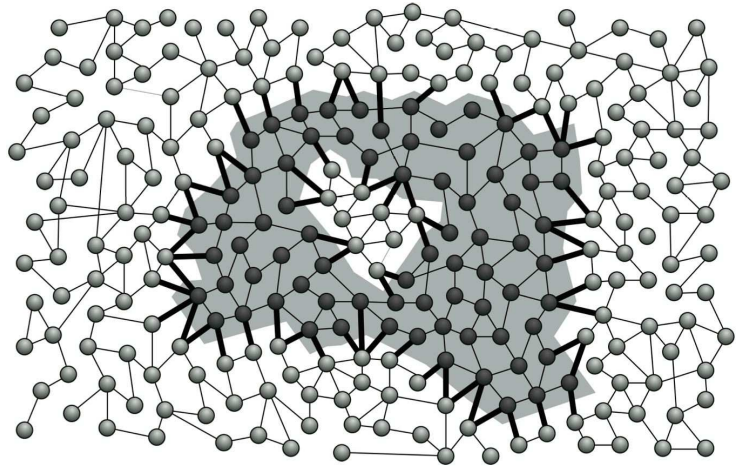}
\caption{A general graph with a distinguished region $I\subset L$ (gray area, dark gray oscillators). Lines indicate the edge set, the number of bold lines is the surface area $s(I)$.\label{boebbel}}
\end{center}
\end{figure}

In this paper, we rigorously 
reconsider this question
for a class of systems on general lattices the constituents of which
are infinite-dimensional quantum systems: the quasi-free case of 
harmonic systems on general lattices (see Figure \ref{boebbel}).
This harmonic case is particularly 
transparent, as the entire discussion of state 
properties can be done in terms of the second moments of the 
states. This case has notably 
taken a central role in recent discussions of 
questions of entanglement
scaling in many-body systems, as a class of 
natural physical 
systems where sophisticated questions on the
scaling of entanglement -- in form of, e.g., 
{\it entanglement-area relationships} 
\cite{Harmonic,Area,Area2} and other issues
\cite{Botero,Frustrated,Frustrated2,Iran,Dynamics} -- 
are relatively accessible.

\begin{figure}
\begin{center}
\includegraphics[width=0.95\textwidth]{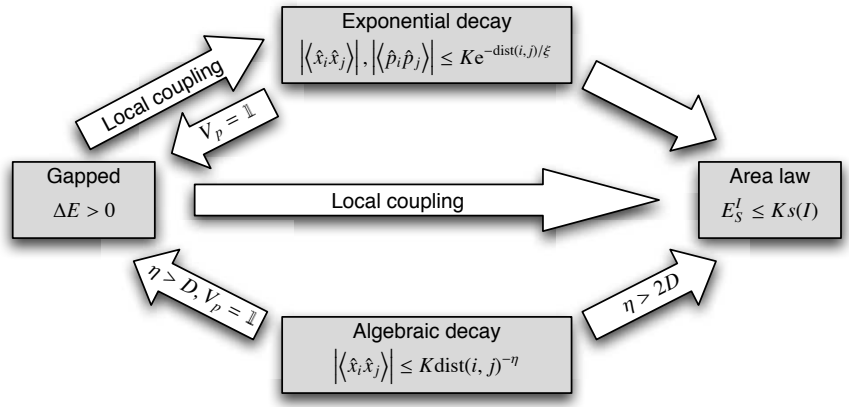}
\caption{Schematic summary of obtained
results.\label{results}}
\end{center}
\end{figure}

The general lattices are not assumed to be
necessarily translationally invariant, and the underlying graphs
may have any spatial dimension. 
Harmonic systems, needless to say, have been considered
many times 
before, in particular the case of nearest-neighbor interaction
\cite{Jaffe}: Here, 
we investigate in a very comprehensive manner the 
implications of a gap to the two-point correlation
functions and the converse direction 
for all harmonic lattice systems 
with local Hamiltonians reflecting general
couplings in position as well as in 
momentum, both for the ground state 
and Gibbs states (non-zero temperature).
Notably,
these findings enable us to formulate an entanglement-area
law for regions of arbitrary shape within the considered context
of harmonic systems on general graphs.  
Such harmonic systems model discrete versions of Klein Gordon fields,
vibrational modes of crystal lattices or of ions in a
trap, or serve as 
approximations to not strictly harmonic systems. 
In contrast to, for example
Ref.\ \cite{Hastings}, we will not rely on 
using variants of Lieb-Robinson \cite{LiebRobinson} 
bounds on the group velocity to prove the validity
of our bounds. Instead 
we will make use of generalizations of
the methods introduced in Ref.\ \cite{Benzi} to the case 
of general lattices (see Figure \ref{results}).

\begin{itemize}

\item We briefly introduce the notion of {\it quasi-free states
as ground or thermal states of harmonic Hamiltonians
on generic lattices} described by graphs, with coupling
in position and momentum coordinates.

\item For {\it local harmonic Hamiltonians on generic lattices} 
we prove that whenever the system is gapped, exponential
decay of the correlations in the {\it ground state}
with the canonical coordinates follows (Theorem 1).

\item Conversely, for systems (not necessarily locally) coupled in position, we demonstrate
that for {\it sufficiently fast decay of the correlations}, the 
existence of a spectral gap can be deduced. This includes
sufficiently strongly algebraically decaying correlations 
(Theorem 2,3).

\item These findings give rise to a proven {\it equivalence
of exponentially decaying couplings and the 
existence of a spectral gap} for systems coupled in 
position on generic lattices (Corollary 1).

\item We prove that for gapped harmonic lattice systems on generic
lattices in {\it Gibbs states} (thermal states), exponential
decay of the correlations follows (Theorem 4). 

\item We give a connection \cite{Mike,Area,Area2} 
to entanglement properties in many-body systems,
in this instance of harmonic lattice systems. We show that 
similarly to Ref.\ \cite{Area2}, the {\it entropy of a subsystem
of a gapped lattice system 
is bounded from above by a quantity linear in the boundary
area} of the distinguished subsystem, which may be of
arbitrary shape (Theorem 5). This generalizes the
{\it area theorem} 
of Refs.\ \cite{Area,Area2} 
to systems on general lattices.

\item We finally discuss several examples in detail,
such as instances of {\it disordered systems} or {\it thermal
states of rotating wave Hamiltonians}. 
We also investigate a case of {\it non-local algebraically 
decaying interactions}.

\end{itemize}
Note that simultaneously, similar findings were published in
Ref.\ \cite{Private}, where in contrast to this paper,
emphasis was put on ground state properties of
translationally invariant harmonic 
systems on cubic lattices, but where
in turn, non-local interactions and critical cases
were considered in much more detail.

\section{Considered models and main results}

\subsection{Harmonic systems on general lattices}

We consider quadratic bosonic Hamiltonians on generic lattices. 
This means that the lattice is characterized by
general graphs \cite{GraphTheory}. 
The set $L$ of vertices with cardinality $|L|$
is associated with the set of sites in 
the lattice, each of which corresponding to 
a bosonic degree of freedom. The simple 
graph $G=(L,E)$ is 
identified with the lattice, where $E$ is the edge set
or, equivalently, the adjacency matrix 
characterizing a neighborhood relation between 
physical systems. In a cubic lattice, say, $G$ will
represent just this general 
lattice, regardless of the interaction,
which may be finite-ranged beyond 
nearest or next-to-nearest
neighbor interactions. 

We will consider paths in the sense of standard graph theory, as sequence of 
vertices connecting a start and an end vertex, each of which are connected by an
edge. For two vertices $i,j\in L$, 
the integer 
$\,\text{dist}(i,j)$ denotes the 
graph-theoretical distance, so the length
of the shortest path connecting $i$ and $j$ \cite{GraphTheory}, with respect to the edge set $E$. 

\begin{itemize}
\item[(i)]
The Hamiltonian is assumed to be of the
form
\begin{equation}
	\label{ham}
	\hat{H}=
		\hat p^T V_p \hat p + \hat x^T V_x \hat x.
\end{equation}
where $V_x,V_p\in \rr^{|L|\times |L|}$ are positive matrices,
with $\hat x^T=(\hat x_1,...,\hat x_{|L|})$ and
$\hat p^T=(\hat p_1,...,\hat p_{|L|})$ being the canonical coordinates
satisfying the canonical commutation relations. Hence, we
allow for any general coupling in position and momentum 
coordinates (but no coupling involving both simultaneously).
We write $(V_{x,p})_{i,j}$, $i,j=1,...,|L|$, 
to label the element of the  
matrices $V_{x,p}$ that belongs to the coupling  
between
two vertices $i,j$, respectively.
For simplicity of notation, these couplings in position
and momentum can be subsumed into the tuple
$C=(G,V_x, V_p)$. Whenever two sites $i,j\in L$
are coupled in position, the element $(V_x)_{i,j}$ will 
be non-zero, and similarly for the momentum
coordinates. The range of the interaction is 
taken with respect to the metric $\text{dist}(\cdot,\cdot)$, 
such that $V_x$ and $V_p$ inherit a neighborhood
relation.
\end{itemize}
Note that $V_x$ and $V_p$ can in turn be conceived
as adjacency matrices of a weighted graph with the
same vertex set $L$, but a different edge set.
The above 
description is similar to the assessment of the correlation function for 
generic spin systems  in Refs.\ \cite{Hastings,Koma}.
Note that such bosonic 
harmonic lattice systems resemble to some extend 
the concept of graph states for 
spin or qubit systems in the sense of 
Refs.\ \cite{Graphs,Graphs2},
compare also Refs.\ \cite{Area2,Frustrated,Iran}.

The dimension of the underlying lattice $G$
can take any positive value in the
following sense:  
We may define a sphere $S_r(i)$ for some $i\in L$, centered at site
$i$ with integer radius $r$ as
\begin{equation}
	S_r(i) : = \left\{
	l\in L: \text{dist}(l,i)=r
	\right\}.
\end{equation}
Then there exists a smallest $d>0$ 
of the lattice, notably not necessarily integer \cite{Koma},
such that for all $r\in \nn$,
\begin{equation}	
\label{dimension}
	\sup_{i\in L} |S_r(i)| \leq c r^{d-1}
\end{equation}	
for some $c>0$. This number $d$ is taken as the dimension of the lattice. Note that for cubic lattices this dimension coincides
with the natural underlying spatial
dimension, and the above expression is the
same for open and periodic boundary conditions.
Similarly we can define a ball centered at 
site $i\in L$ with radius $r$ as
\begin{equation}
	B_r(i) := \left\{l \in L : \text{dist}(l,i)\le r \right\}
\end{equation}
and can express the maximal volume 
\begin{equation}
	v_{d,r}: = \sup_{i\in L} |B_r(i)| \leq c \sum_{j=1}^{r} j^{d-1}
\end{equation}
of a ball with radius $r$ in the graph theoretical sense.
For a consideration of the 
dimension of general graphs and
the discussion of self-similarity in this context, 
see, e.g., Ref.\ 
\cite{GeneralGraphs}. For our purposes, the dimension will
only enter the bounds through the volume of a ball of some radius.

%
Concerning the locality, we assume the following: 
\begin{itemize}
\item[(ii)]
Most Hamiltonians that we consider are local, corresponding to
a  finite-ranged interaction, if not otherwise specified. 
This means that there exists an $m\in \nn$, 
such that for all $i\in L$
\begin{equation}
	(V_x)_{i,j}=0,\,\, (V_p)_{i,j}=0
\end{equation}
for all $j\in L$ for which
\begin{equation}
	\text{dist}  (i,j) >m/2.
\end{equation}

Note that this in turn means that 
the Hamiltonian is of the form as in Eq.\ (\ref{ham})
and can be written as
\begin{equation}
	\hat H=\sum_{\substack{ X\subset  L,\\ 
	\text{diam}(X)\le m/2}} \hat h_X,
\end{equation}
$X\subset L$ being subsets the diameter of which
satisfy
\begin{equation}
	\text{diam}(X)=\sup_{i,j\in X}\text{dist}(i,j)\le m/2.
\end{equation}
\end{itemize}

It will turn out to be convenient to collect the $|L|$ 
conjugate pairs of canonical coordinates in a vector
${\hat{r}}^T=(\hat{x}_1,...,\hat{x}_{|L|},\hat{p}_1,...,\hat{p}_{|L|})$,
the entries of which satisfying the 
canonical commutation relations (CCR), giving rise to a 
symplectic scalar product \cite{Survey}.
The Hamiltonian
\begin{equation}
	\hat{H}= \hat{r}^T\left(\begin{array}{cc}
	{V}_x & 0\\
	0 & {V}_p
\end{array}\right) \hat{r},
\end{equation}
can now be brought into diagonal form
by means of a symplectic transformation, so by means of 
linear transformations $S\in Sp(2|L|,\rr)$
preserving the symplectic form \cite{Survey}, 
defined by the skew-symmetric 
symplectic matrix
\begin{eqnarray}
\Sigma=\left(\begin{array}{cc}
0 & \id\\
-\id & 0
\end{array}\right).
\end{eqnarray} 
Let $O\in O(|L|)$ be the orthogonal matrix that 
brings $V_x^{1/2}V_pV_x^{1/2}$ into
diagonal form, 
$O^TV_x^{1/2}V_pV_x^{1/2}O=:D=\text{diag}(d_1,\cdots,d_{|L|})$. 
Then
\begin{equation}
	S=\left(\begin{array}{cc}
	V_x^{-1/2}O&0\\
	0&V_x^{1/2}O
	\end{array}\right)\in Sp(2|L|,\rr)
\end{equation}
and with $\hat{r}=S\hat{r}'$, the Hamiltonian finally takes the form
\begin{equation}
\label{diagBasis}
	\hat{H}
	=2\sum_{i=1}^{|L|} d_i^{1/2}
	\left(\hat{a}_i^\dagger \hat{a}_i+1/2 \right),\,\,\,
	\hat{a}_i:=\frac{\hat{x}'_i+\mi d_i^{1/2}\hat{p}'_i}{
	(2 d_i^{1/2})^{1/2}}.
\end{equation}
The ground state energy is thus given by
\begin{equation}
	E_0=\sum_{i=1}^{|L|}d_i^{1/2}
	=\text{tr}[(V_x^{1/2}V_pV_x^{1/2})^{1/2}]=\text{tr}[( 	{V}_x {V}_p)^{1/2}].
\end{equation}
The energy of the first excited state is given by
\begin{eqnarray}
	E_1 &=& E_0+ \Delta E ,\\
	\Delta E 
	&:= & 2 \lambda^{1/2}_{\text{min}}(V_x^{1/2} V_p V_x^{1/2})=2 
	\lambda^{1/2}_{\text{min}}(V_x V_p ),
\end{eqnarray}
where we denote by $\lambda_{\text{min}}(A)$ the minimum eigenvalue of a matrix $A$.
The corresponding ground state of such a Hamiltonian is a 
quasi-free (Gaussian) state \cite{Survey}, 
completely characterized by the first and second moments.
Quasi-free bosonic or Gaussian state means that the
characteristic function -- expectation value of the
Weyl displacement operator -- is a Gaussian function in 
state space. Equivalently, a state is Gaussian if its
Wigner function is a Gaussian function in state space.
The second moments
can be embodied in the covariance matrix with respect to the
ground state
\begin{eqnarray}
	\gamma_{i,j}' &=&\left\langle \{(\hat r_i'-\langle \hat r_i'\rangle),(\hat r_j'-\langle 	\hat r_j'\rangle)\}_+\right\rangle\nonumber \\
	&=&\langle \hat r_i'\hat r_j'\rangle+\langle \hat r_j'\hat r_i'\rangle.
\end{eqnarray}
Here, first moments vanish as the ground state is the vacuum. 
We find
\begin{equation}
	\gamma'={D}^{1/2} 
	\oplus {D}^{-1/2},
\end{equation}
and in the original coordinates we have
\begin{eqnarray}\label{posg}
	\gamma &= & S\gamma' S^T \\
	 &=& \left(
	 V_x^{-1/2} (V_x^{1/2} V_p V_x^{1/2})^{1/2} V_x^{-1/2}
	 \right)
	\oplus
	\left(
	 V_x^{1/2} (V_x^{1/2} V_p V_x^{1/2})^{-1/2} V_x^{1/2}
	 \right).\nonumber
\end{eqnarray}
This expression simplifies significantly in case of commuting
$V_x$ and $V_p$. Often, $V_p=\id$, in which case
\begin{equation}
	\gamma = V_x^{-1/2} \oplus V_x^{1/2},
\end{equation}	
see Ref.\
\cite{Harmonic}. 
The language chosen here is the one used also in the
assessment of entanglement properties of 
harmonic chains \cite{Harmonic,Botero}
or more general harmonic systems on lattices 
\cite{Area,Area2,Survey,Frustrated,Frustrated2,Maciek}, 
mildly generalized to coupling in both 
position and momentum and to general lattices. 

%

\subsection{Exponential decay of correlations in ground states}

The first result links the spectral gap of the Hamiltonian to the 
exponentially decaying correlation functions of the system. Note that 
the bound on the right hand side depends only on the dimension of the underlying
graph, the gap, the parameter characterizing the range of the interactions,
and bounds to the coupling strength. 

\begin{theorem}[Exponentially decaying correlation functions]\label{theorem1}
	Consider a Hamiltonian on a general lattice of dimension $d$ 
	with a coupling 
	$C=(G,V_x,V_p)$ of finite range $m$
	as defined above. 
	Then for all $i,j\in L$ with $\text{dist}(i,j)\ge m$ the 
	ground state satisfies
	\begin{eqnarray}
		| \langle \hat x_i \hat x_j\rangle| 
		&\leq&  K \| V_p \|
		 \exp\left[-\text{dist}(i,j)/\xi\right],\\
		| \langle \hat p_i \hat p_j\rangle |
		&\leq& K\| V_x \|
		 \exp\left[ -\text{dist}(i,j)/\xi\right],
	\end{eqnarray}
	where 
	\begin{equation}
		K  := 
		\frac{\| V_xV_p \|^{1/2}  v_{d,m/2}}{(\Delta E/2)^2},\;\;\;
	\xi : =
	\frac{2m}{\log\left(\frac{\|V_xV_p\|}{\|V_xV_p\|-(\Delta E/2)^2}\right)},
	\end{equation}
	where $\|\cdot\|$ denotes the operator norm and $\Delta E=2\lambda_{\text{min}}^{1/2}(V_xV_p)$ is the energy difference between the first excited state and the ground state.
\end{theorem}
 
 Note that if the energy is expressed in units of 
 $\| V_x V_p\|$, then the correlation length depends
 only on the gap and the range of the interactions.  

\subsection{Spectral 
gap from algebraically decaying correlation functions}

In this subsection, we consider an instance of the converse
direction, compare Ref.\ \cite{LSM}: 
we assume that the system has algebraically
decaying correlation functions in the position coordinates
and assume that we have a -- not necessarily local -- coupling 
in these coordinates and $V_p=\id$,
then we can conclude that the system must be gapped if the
decay is sufficiently strong. 
This is made more rigorous in the subsequent theorem.

\begin{theorem}[Spectral gap from algebraic
decay] Consider a sequence of couplings
	  $C^{(n)}=(G^{(n)},V^{(n)},\id)$, $n\in \nn$,
	on general lattices
	   $G^{(n)} = (L^{(n)},E^{(n)})$ of 
	   dimension $d^{(n)}$. Let
	   $K\ge 0$, $K_0\ge 0$, $\eta>\sup d^{(n)}=:d$, and $c:=\sup c^{(n)}<\infty$ as defined in Eq.\ (\ref{dimension}).  If the
	   ground states satisfy
\begin{eqnarray} 
\label{xcorr}
|\langle \hat x^{(n)}_i \hat x^{(n)}_j\rangle|
		&\leq&  
		\begin{cases}
		K_0 &\text{ for } i=j,\\
		K \text{dist}(i,j)^{-\eta} &\text{ for } i\ne j,
		\end{cases}
\end{eqnarray}   
for all $i,j\in L^{(n)}$ and for all $n$, then the energy
difference between the first excited and the ground state satisfies
\begin{equation}
\inf \Delta E^{(n)}=:\Delta E\ge\frac{2}{K_0+cK\zeta(1+\eta-d)}>0,
\end{equation}
where $\zeta$ is the Riemann zeta function.
\end{theorem}

Note that, if in addition to the assumptions of the above theorem the coupling $V^{(n)}$ is (i) of finite range $m$ for all $n$,
and (iia) the momentum correlations decay as in Eqs.\ (\ref{xcorr}) or alternatively, (iib) the coupling
has a finite coupling strength, $\sup\|V^{(n)}\|<\infty$,\footnote{Note that (iib) follows from (iia) using the same methods as in the proof for Theorem 2.}
it immediately follows from Theorem 1 that in fact the correlations decay exponentially. This is an interesting observation, showing
that in a sense the class of harmonic lattice systems is not
rich enough to -- roughly speaking -- 
show all ``types of decays'' of the correlations. 

\subsection{Equivalence of spectral 
gap and exponentially decaying correlation functions}
For exponentially decaying correlations we can always conclude that the 
system is gapped. This statement can be formulated as follows.
\begin{theorem}[Spectral gap from exponential decay] Let $K\ge 0$ and $\xi\ge 0$, and 
consider a sequence of couplings
	  $C^{(n)}=(G^{(n)},V^{(n)},\id)$, $n\in \nn$,
	on general lattices
	   $G^{(n)} = (L^{(n)},E^{(n)})$ of 
	   dimension $d^{(n)}$, $d:=\sup d^{(n)}<\infty$, and 
	   $c:=\sup c^{(n)}<\infty$. If the ground state satisfies 
\begin{equation} 
|\langle \hat x^{(n)}_i \hat x^{(n)}_j\rangle|
		\leq  K 
		 \exp\left[-\text{dist}(i,j)/\xi\right]
\end{equation}   
for all $i,j\in L^{(n)}$ and all $n$, then
the energy
difference between the first excited and the ground state satisfies
\begin{equation}
\inf \Delta E^{(n)}=:\Delta E\ge\frac{2}{K(1+cLi_{1-d}(\me^{-1/\xi}))}>0,
\end{equation}
where $Li_{1-d}$ is the polylogarithm of degree $1-d$.
\end{theorem}

Together with Theorem 1 this establishes
for locally coupled systems 
the following equivalence.

\begin{corollary}[Equivalence of spectral gap and exponentially decaying correlations]
Consider the sequence of couplings
	  $C^{(n)}=(G^{(n)},V^{(n)},\id)$, $n\in \nn$, of finite range $m$
	on general lattices
	   $G^{(n)} = (L^{(n)},E^{(n)})$ of dimension $d^{(n)}$ with $d:=\sup d^{(n)}<\infty$, and 
	   $c:=\sup c^{(n)}<\infty$.
	    Then the following statements are equivalent.
\begin{itemize}
\item[(i)] There exist constants $K,\xi>0$ such that 
both correlations with respect to the ground state
satisfy
\begin{equation} 
	|\langle \hat x^{(n)}_i \hat x^{(n)}_j\rangle|, 
	|\langle \hat p^{(n)}_i \hat p^{(n)}_j\rangle|
		\leq  K 
		 \exp\left[-\text{dist}(i,j)/\xi\right]
\end{equation}  
for all $i,j\in L^{(n)}$ and all $n$. 
\item[(ii)] The energy difference between the first excited state and the ground state satisfies
\begin{equation}
\inf \Delta E^{(n)}=:\Delta E>0
\end{equation}
and the coupling is of finite strength, $ \sup\|V^{(n)}\|<\infty$.
\end{itemize}
\end{corollary}

This result establishes that in local 
harmonic systems on generic
lattices, being non-critical in the sense of being 
gapped is equivalent with finding
exponentially decaying correlation functions. 

\subsection{Expontial decay in Gibbs states}

In this subsection, we
consider thermal
Gibbs states \cite{Survey} 
corresponding to some 
temperature $T>0$, 
\begin{equation}
\varrho(T)=\frac{\exp(-\hat{H}/T)}{\text{tr}[\exp(-\hat{H}/T)]}.
\end{equation}
Such states are again
quasi-free (Gaussian) 
states and can thus also be uniquely
characterized by their
covariance matrices, appropriately modified for this
case of non-zero temperature.
For a study of two-point correlations in Fermi-systems
at non-zero temperature see also 
Ref.\ \cite{HastingsFermion}.
In the diagonal basis (\ref{diagBasis}), it takes the
form
\begin{eqnarray}
\gamma(T)'&=&D^{1/2}\left[\id+2\bigl(\exp(2 D^{1/2}/T)-\id\bigr)^{-1}\right]\nonumber\\
& &\hspace{1cm}\oplus\;
D^{-1/2}\left[\id+2\bigl(\exp(2 D^{1/2}/T)-\id\bigr)^{-1}\right]
\end{eqnarray}
and in original coordinates we have
\begin{eqnarray}
\gamma(T)=S\gamma(T)'S^T&=&
\gamma(0)+
V_x^{-1/2}G
\bigl(V_x^{1/2}V_pV_x^{1/2}\bigr)^{1/2}V_x^{-1/2}\nonumber\\
& &\hspace{2cm}\oplus\;
V_x^{1/2}\bigl(V_x^{1/2}V_pV_x^{1/2}\bigr)^{-1/2}GV_x^{1/2},\\
G&:=&2\left(\exp\left(2 (V_x^{1/2}V_pV_x^{1/2})^{1/2}/T\right)-\id\right)^{-1}.
\end{eqnarray}
In the following we will now assume that matrices $V_x$ and
$V_p$ commute. This yields a simplified covariance matrix
\begin{eqnarray}
\gamma(T)&=&
\gamma(0)+\bigl(V_x^{-1/2}V_p^{1/2}G\bigr)\oplus\bigl(V_x^{1/2}V_p^{-1/2}G\bigr),\\
G&=&2\left(\exp\bigl(2 (V_xV_p)^{1/2}/T\bigr)-\id\right)^{-1}.
\end{eqnarray}
Furthermore, we require the following assumption
on the lattice. This assumption is very similar to the ones in 
Ref.\ \cite{Hastings},
and is 
satisfied for a large class of natural lattice systems.



\begin{assumption}[Lattice structure] Consider a Hamiltonian on a general lattice $G=(L,E)$ with couplings $C=(G,V_x,V_p)$ as in Theorem 1.
Then, it is assumed that there exist constants $l_0\geq 0$
and $\nu >0$ such that 
\begin{equation}
\sum_{k\in L}\exp(-\mu\text{dist}(i,k))\exp(-\mu\text{dist}(k,j))
\leq
l_0\exp(-\nu\text{dist}(i,j)),
\end{equation}
for all $i,j\in L$ and for 
\begin{equation}
\mu:=
\frac{\log\left(\frac{\|V_xV_p\|}{\|V_xV_p\|-(\Delta E/2)^2}\right)}{m}.
\end{equation}
\end{assumption}

For example, for a cubic lattice in $d$ dimensions, we have that 
$L=[1,...,n]^{\times d}$, 
and $\text{dist}(i,j)=\sum_{\delta=1}^d 
|i_\delta-j_\delta|$. 
Therefore,
we arrive at
\begin{eqnarray}
	\sum_{k\in L}\me^{-\mu\text{dist}(i,k)}\me^{-\mu\text{dist}(k,j)}=
	\prod_{\delta=1}^d\sum_{k_\delta=1}^n\me^{-\mu|i_\delta-k_\delta |}\me^{-\mu|k_\delta-	j_\delta|}\nonumber\\
	=\prod_{\delta=1}^d\left(
	\me^{-\mu |i_\delta-k_\delta|}\left(|i_\delta-j_\delta|+\frac{\me^{2\mu}+1}{\me^{2
	\mu}-1}\right)-\frac{\me^{-\mu(i_\delta+j_\delta-2)}+\me^{-\mu(2n-i_\delta-j_\delta)}}
	{\me^{2\mu}-1}\right)
	\nonumber \\
	\leq
	\left(\frac{2}{\mu\me}+\frac{\me^{2\mu}+1}{\me^{2\mu}-1}\right)^d
	\exp(-\mu\text{dist}(i,j)/2),
\end{eqnarray}
i.e., in this cubic case, we have $\nu=\mu/2$.

Under the previous assumption, one can arrive at the
subsequent statement on two-point correlations in Gibbs
states. Note that it is not merely 
trivially true that the correlations
are shorter-ranged in thermal states: there
are examples of Hamiltonians where Gibbs states at higher
temperatures have longer-ranged correlations than at
zero temperature, see Example \ref{thermalexample}.

\begin{theorem}[Correlations in systems at finite temperature]
Consider a finite-ranged Hamiltonian corresponding to
$C=(G,V_x,V_p)$, $[V_x,V_p]=0$, 
on a general lattice 
	$G=(L,E)$   
	with assumptions as in
	Theorem \ref{theorem1} and equipped with 
	Assumption 1. 
	Then, for $\text{dist}(i,j)\geq m$, the Gibbs state with respect to 
	some temperature $T$ satisfies
	\begin{eqnarray}
		| \langle \hat x_i \hat x_j\rangle| 
		&\leq&  K(T) \| V_p \|
		 \exp\left[-\text{dist}(i,j)/\xi\right],\\
		| \langle \hat p_i \hat p_j\rangle |
		&\leq& K(T)\| V_x \|
		 \exp\left[ -\text{dist}(i,j)/\xi\right],
	\end{eqnarray}
where
	\begin{equation}
		K(T):= \frac{\|V_xV_p\|^{1/2}v_{d,m/2}}{(\Delta E/2)^2}
		\left(
		    1+\frac{4l_0\|V_xV_p\|/(\Delta E/2)^2}{
		    \exp\left(	
		    \frac{\Delta E}{T}
		    \left({{1-\frac{(\Delta E/2)^2}{4\|V_xV_p\|}}}\right)^{1/2}
		    \right)-1}\right),
	\end{equation}
	\begin{equation}
	\xi:=\max\left\{
	\frac{2}{\nu},\frac{2m}{\log\left(\frac{\|V_xV_p\|}{\|V_xV_p\|-(\Delta E/2)^2}\right)}
	\right\}.
	\end{equation}
\end{theorem}

This theorem ends the list of main statements on 
the relation between the system being gapped and
the decay of the two-point correlation functions with respect
to the ground state and thermal states.
In the next section we will study the implications of the
above findings on the entanglement scaling of distinguished
regions of general lattices

\subsection{Entanglement scaling in ground states of harmonic systems on general lattices}
The above  statements on the decay of correlation
functions have immediate implications on the scaling 
of entanglement. More specifically, in a lattice system
we may distinguish a certain part $I\subset L$ and ask for the 
degree of entanglement between the degrees of freedom
of this region and the rest of the lattice. This question
is that of the scaling of the
{\it geometric entropy of this region}.
This question goes back to seminal numerical  investigations
in Refs.\ \cite{AreaOld1,AreaOld2} (see also 
 Ref.\ \cite{AreaOld3}), suggesting a linear relationship
 between 
 the geometric entropy
 and the boundary area of a distinguished region.
 In turn, 
in Refs.\ \cite{Area,Area2} a rigorous 
relationship between the 
boundary area and the entropy of the region 
in cubic harmonic lattice systems
has first been analytically established.

In fact, in the light of the above findings, one 
may infer the validity of such an 
'area theorem' for general lattices, not only for cubic lattices,
exploiting exactly the same methods of proof. Defining the surface
area of a distinguished region $I$ of the whole lattice
as 
\begin{equation}
	s(I):=\sum_{i\in L\backslash I}\sum_{\substack{j\in I\\ \text{dist}(i,j)=1}}\!\!\!\!\! 1,
\end{equation}
see Figure \ref{boebbel},
one arrives at a
bound to the von-Neumann entropy,
\begin{equation}
E_S^I=S(\varrho_I)=-\text{tr}\left[\varrho_I\log_2(\varrho_I)\right],
\end{equation}
of the reduced ground state,
$\varrho_I=\text{tr}_{L\backslash I}\left[\varrho\right]$, with respect to the distinguished region.
 
\begin{theorem}[Entanglement-area law for general lattices]
Consider a general lattice $G=(L,E)$ of dimension $d$ equipped with
a coupling $(G,V,\id)$. If the coupling
is of finite range $m$, the entropy of entanglement satisfies
\begin{equation}
	E_{S}^I\leq \frac{4\| V\| c^2 Li_{1-2d}
	(\me^{-1/\xi})}{\log(2)(\Delta E/2)^2}s(I), 
\end{equation}
where $Li_{1-2d}$ is the polylogarithm of degree $1-2d$, $\Delta E$ the energy gap above the ground state, $s(I)$ the surface area of $I$, $c$ as defined in Eq.~(\ref{dimension}), and 
\begin{equation}
	\xi:=
	\frac{m}{\log\left(\frac{\| V\|}{\| V\|-(\Delta E/2)^2}\right)}.
\end{equation}
\end{theorem}
In this form, the contribution of a geometrical 
factor, as well as one originating from the correlation
length, is very transparent.
Note that, using Theorem 4 and the bounds derived in Ref.\ \cite{Area2}, one arrives at a similar bound for
the distillable entanglement of thermal states.

Indeed, 
this theorem covers the entanglement-area relationship
in all generality for gapped harmonic models on generic
lattices. 
In particular,  for cubic regions 
with volume $L^d$ in $d$ dimensions
this means
that the geometric entropy is bounded by 
expressions linear in $L^{d-1}$. For critical
fermionic quasi-free systems, in turn, 
there are instances where one
finds a behavior of $L^{d-1}\log L$ 
for the geometric entropy \cite{Gioev,Wolf}.
In contrast, there is numerical evidence 
that for bosonic harmonic systems, even
in the critical case, 
the validity of the entanglement-area 
relationship of $L^{d-1}$ is preserved 
\cite{Julian,Scholl}.
To strictly prove or refute 
this relationship in this critical case
constitutes one of the intriguing open 
problems in the field.

\section{Proofs}

This section will contain the proofs of the statements
made before. We will make
extended use of the bandedness of the interaction matrix
in the sense of the metric $\text{dist}(\cdot,\cdot)$. The main
ingredient will be polynomial approximations to 
matrix functions of matrices. We generalize a statement
of Ref.\ \cite{Benzi} to the case of general lattices.
This extends the generalization put forth in Ref.\ 
\cite{Area2} for general cubic lattices. 
Note that for statements of the type following, random walk
methods would similarly be suitable. In particular,
for the case of nearest-neighbor interaction, this analysis 
has been done, see, e.g., Ref.\ \cite{Jaffe} for a 
comprehensive introduction.

In the following we will be needing the subsequent 
lemma which is concerned with
the range of matrices in the sense of $\text{dist}(\cdot,\cdot)$.
Note that the matrix power is taken in the sense of the
ordinary matrix power.\\

\begin{lemma}[Range of matrix powers]
Let the $|L|\times|L|$ matrix $A$ be a coupling matrix on a general lattice $G=(L,E)$. If $A$ is of finite range $m$, i.e., $A_{i,j}=0$ 
for
$\,\text{dist}(i,j)>m/2$, $m\in\nn$, then $A^n$, $n\in\nn$, is of range $nm$.
\end{lemma}

\proof
This can be proven by induction over $n$.
For $n=1$ we have $A^n=A$ is of range $nm$. 
Now suppose $A^n$ has range $nm$, i.e.,
we have $(A^n)_{i,j}=0$ for $\,\text{dist}(i,j)>nm/2$. Now,
\begin{equation}
\left(A^{n+1}\right)_{i,j}
=
\sum_{k\in L}\left(A^n\right)_{i,k}A_{k,j}.
\end{equation}
Suppose $\,\text{dist}(i,j)> {m(n+1)}/{2}$. It follows that
for $\,\text{dist}(i,k) > nm/2$ 
we have $(A^n)_{i,k}=0$ as $A^n$ is of range $nm$. 
For
$\,\text{dist}(i,k)\le nm/2$ we have 
\begin{equation}
m(n+1)/2< \,
\text{dist}(i,j)\le \,\text{dist}(i,k)+\,\text{dist}(k,j)
\le nm/2+\,\text{dist}(k,j),
\end{equation}
i.e., $\,\text{dist}(k,j)> m/2$, i.e., $A_{k,j}=0$.
In turn, 
for $\,\text{dist}(k,j) \le m/2$ we have 
\begin{equation}
	m(n+1)/2< \,
	\text{dist}(i,j)\le \,\text{dist}(i,k)+\,\text{dist}(k,j)
	\le \,\text{dist}(i,k)+m/2,
\end{equation}
i.e., $\,\text{dist}(i,k)> nm/2$, i.e., 
$(A^n)_{i,k}=0$ as $A^n$ is of range $nm$.
Finally, we hence arrive at $(A^{(n+1)})_{i,j}=0$ 
for $\text{dist}(i,j)>m(n+1)/2$, which concludes the proof.\qed

\subsection{Proof of Theorem 1}

From the covariance 
matrix $\gamma$, Eq.\ (\ref{posg}), we have
\begin{eqnarray}
\langle\hat{p}_i\hat{p}_j\rangle&=&
	\left(
	 V_x^{1/2} (V_x^{1/2} V_p V_x^{1/2})^{-1/2} V_x^{1/2}
	 \right)_{i,j}, \\
	\langle\hat{x}_i\hat{x}_j\rangle&=&
	\left(
	 V_x^{-1/2} (V_x^{1/2} V_p V_x^{1/2})^{1/2} V_x^{-1/2}
	 \right)_{i,j}\nonumber\\
	&=&
	\left(
	 V_p^{1/2} (V_p^{1/2} V_x V_p^{1/2})^{-1/2} V_p^{1/2}
	 \right)_{i,j}.
\end{eqnarray}
We can now compute the matrix $(V_x^{1/2} V_p V_x^{1/2})^{-1/2}$
and similarly $(V_p^{1/2} V_x V_p^{1/2})^{-1/2}$
using the power-series expansion of the square root, which is valid for $|x|<1$,
\begin{equation}
\label{squareroot}
(1-x)^{-1/2}=1+\sum_{k=1}^\infty a_k x^k,
\end{equation}
where the coefficients $a_k$ are bounded by $a_k \leq 1$. 
\begin{eqnarray}
\left(V_x^{1/2}V_pV_x^{1/2}\right)^{-1/2}
=\| V_xV_p \|^{-1/2}\left(\id-(\id-V_x^{1/2}V_pV_x^{1/2}/\| V_xV_p \|)\right)^{-1/2}\nonumber \\
=
\| V_xV_p \|^{-1/2}\left(\id + \sum_{k=1}^\infty a_k O\left(\id- D/\| V_xV_p \| \right)^kO^T\right).
\end{eqnarray}
Now,
\begin{eqnarray}
	D=O^TV_x^{1/2}V_pV_x^{1/2}O
	=\left(V_x^{-1/2}O\right)^{-1}V_pV_x\left(V_x^{-1/2}O\right),
\end{eqnarray}
i.e.,
\begin{eqnarray}
\left(V_x^{1/2}V_pV_x^{1/2}\right)^{-1/2}
&=&\| V_xV_p \|^{-1/2}V_x^{1/2}W_{p,x}V_x^{-1/2},\\
W_{p,x}&:=&
\left(\id + \sum_{k=1}^\infty a_k\left(\id-\left(V_pV_x\right)/\| 
V_xV_p \| \right)^k\right)
\end{eqnarray}
and similarly
\begin{equation}
\left(V_p^{1/2}V_xV_p^{1/2}\right)^{-1/2}
=\| V_xV_p \|^{-1/2}V_p^{1/2}W_{x,p}
V_p^{-1/2}.
\end{equation}
Note that clearly, $(V_x^{1/2} V_p V_x^{1/2})^{-1/2}$
is a positive matrix, whereas $W_{p,x}$ is in general
not a symmetric matrix (and the matrix power is the $k$-fold
concatenation of matrix multiplication).
Thus, the correlation functions take the form
\begin{eqnarray}
\langle\hat{x}_i\hat{x}_j\rangle
&=&
\| V_xV_p \|^{-1/2}
\left(
V_pW_{p,x}\right)_{i,j},\\
\langle\hat{p}_i\hat{p}_j\rangle
&=&
\| V_xV_p \|^{-1/2}
\left(
V_xW_{x,p}\right)_{i,j}.
\end{eqnarray}
Assuming $V_x$ and $V_p$ to be of range $m/2$, i.e.,
$(V_x)_{i,j}=(V_p)_{i,j}=0$ for $\text{dist}(i,j)>m/2$, we have that $V_xV_p$
and $V_pV_x$ are of range $m$.
Note that for a matrix to be of finite range in the sense of the above
Lemma it is not required that it is symmetric.
Hence we can conclude that 
$(V_xV_p)^k$
and $(V_pV_x)^k$ are of range $km$.
Thus
\begin{equation}
\left|\left(W_{p,x}\right)_{i,j}\right|
=
\left|
\delta_{i,j} + \sum_{k\ge \lceil \text{dist}(i,j)/m\rceil} a_k
\left(\left(\id-\left(V_pV_x\right)/\| V_xV_p \|\right)^k\right)_{i,j}\right|.
\end{equation}
For the diagonal terms we have $|(W_{p,x})_{i,i}| \leq \| W_{p,x}\|=\|V_xV_p\|^{1/2}/(\Delta E/2)$
and for the off-diagonal terms we find
\begin{eqnarray}
\left|\left(W_{p,x}\right)_{i,j}\right|
&\leq&
\sum_{k\ge \lceil \text{dist}(i,j)/m\rceil} a_k \left\| \id-\left(V_pV_x\right)/\| V_xV_p \|\right\|^k\nonumber\\
&\leq&
\sum_{k\ge \lceil \text{dist}(i,j)/m\rceil}^\infty \left( 1-\frac{(\Delta E/2)^2}{\| V_xV_p \|}\right)^k.
\end{eqnarray}
The same bound holds for the entries of $W_{x,p}$, i.e., for all $i,j\in L$ we have
\begin{equation}
\left|\left(W_{p,x}\right)_{i,j}\right|,\left|\left(W_{x,p}\right)_{i,j}\right|
\leq
\frac{\| V_xV_p \|}{(\Delta E/2)^2}\left( 1-\frac{(\Delta E/2)^2}{\| V_xV_p \|}\right)^{\text{dist}(i,j)/m}.
\end{equation}
For the correlation function $\langle \hat{x}_i\hat{x}_j\rangle$
this yields
\begin{eqnarray}
|\langle \hat{x}_i\hat{x}_j\rangle|
&\leq&
\frac{\| V_xV_p \|^{1/2}}{(\Delta E/2)^2}
\sum_{k \in L}
\left|(V_p)_{i,k}\right|
\left( 1-\frac{(\Delta E/2)^2}{\| V_xV_p \|}\right)^{\text{dist}(k,j)/m}\nonumber \\
&=&
\frac{\| V_xV_p \|^{1/2}}{(\Delta E/2)^2}
\sum_{\substack{k \in L,\\ \text{dist}(k,i) \leq m/2}}
\left|(V_p)_{i,k}\right|
\left( 1-\frac{(\Delta E/2)^2}{\| V_xV_p \|}\right)^{\text{dist}(k,j)/m}.
\end{eqnarray}
Now $\text{dist}(i,j) \leq \text{dist}(i,k) + \text{dist}(k,j) \leq m/2 + \text{dist}(k,j)$, 
and thus
we find for $\text{dist}(i,j)\ge m$
\begin{eqnarray}
\label{bandedproduct}
|\langle \hat{x}_i\hat{x}_j\rangle|
&\leq&
\frac{\| V_xV_p \|^{1/2}}{(\Delta E/2)^2}
\sum_{\substack{k \in L,\\ \text{dist}(k,i) \leq m/2}}
\left|(V_p)_{i,k}\right|
\left( 1-\frac{(\Delta E/2)^2}{\| V_xV_p \|}\right)^{\text{dist}(i,j)/(2m)}
\nonumber\\
&\leq&
\frac{\| V_xV_p \|^{1/2}}{(\Delta E/2)^2}
\left( 1-\frac{(\Delta E/2)^2}{\| V_xV_p \|}\right)^{\text{dist}(i,j)/(2m)}
\|V_p\|
\sum_{\substack{k \in L,\\ \text{dist}(k,i) \leq m/2}}
\!\!\!\!\!\!\!\!1\;.
\end{eqnarray}
For $|\langle \hat{p}_i\hat{p}_j\rangle|$ one can argue in the
same manner. Noting that the sum on the right hand side is just $|B_{m/2}(i)|$, we finally have for $\text{dist}(i,j)\ge m$
\begin{eqnarray}
|\langle \hat{x}_i\hat{x}_j\rangle|
&\leq&
\frac{\| V_xV_p \|^{1/2}v_{d,m/2}}{(\Delta E/2)^2}\|V_p\|
\left( 1-\frac{(\Delta E/2)^2}{\| V_xV_p \|}\right)^{\text{dist}(i,j)/(2m)},\\
|\langle \hat{p}_i\hat{p}_j\rangle|
&\leq&
\frac{\| V_xV_p \|^{1/2}v_{d,m/2}}{(\Delta E/2)^2}\|V_x\|
\left( 1-\frac{(\Delta E/2)^2}{\| V_xV_p \|}\right)^{\text{dist}(i,j)/(2m)},
\end{eqnarray}
which concludes the proof of Theorem 1.\qed

\subsection{Proof of Theorem 2}

For each $n$, the energy gap is given by
\begin{equation}\label{newgap}
\Delta E^{(n)}=2\lambda_{\text{min}}((V^{(n)})^{1/2})=
\frac{2}{\|(V^{(n)})^{-1/2}\|}.
\end{equation}
The operator norm is bounded from above my the
norm corresponding to the
maximum sum of elements of any row of the matrix, and
hence we may write
\begin{eqnarray}
	\|(V^{(n)})^{-1/2}\|
	&\leq& 
	\sup_{i\in L^{(n)}}
	\sum_{j \in L^{(n)}}|((V^{(n)})^{-1/2})_{i,j}|\\
	&\leq&
	\sup_{i\in L^{(n)}}
	\Biggl(K_0+K\sum_{\substack{j \in L^{(n)}\\ j\ne i}}
	\text{dist}(i,j)^{-\eta}\Biggr)\nonumber
\end{eqnarray}
and hence
\begin{eqnarray}
	\|(V^{(n)})^{-1/2}\|
	&\leq&
	\sup_{i\in L^{(n)}}
	\Biggl(K_0+K\sum_{r=1}^\infty r^{-\eta}\sum_{\text{dist}(i,j)=r}1
	\Biggr),
\end{eqnarray}
and therefore
\begin{eqnarray}
	\|(V^{(n)})^{-1/2}\|
	&\leq&
	\sup_{i\in L^{(n)}}
	\left(K_0+K\sum_{r=1}^\infty r^{-\eta}|S_r(i)|\right)\nonumber\\
	&\leq&
	\left(K_0+Kc^{(n)}\sum_{r=1}^\infty r^{d^{(n)}-1-\eta}\right)\nonumber\\
	&=&K_0+c^{(n)}K\zeta(1+\eta-d^{(n)}),
\end{eqnarray}
where $\zeta$ is the Riemann zeta function, and $c^{(n)}>0$
is defined as in Eq.\ (\ref{dimension}). 
Thus, we can conclude that
\begin{equation}
	\Delta E^{(n)}\ge\frac{2}{K_0+c^{(n)}K\zeta(1+\eta-d^{(n)})}
	\ge\frac{2}{K_0+cK\zeta(1+\eta-d)}>0,
\end{equation}
independent of $n$.
\qed

\subsection{Proof of Theorem 3}

For each $n$, the energy gap is given by
Eq.\ (\ref{newgap}).
Analogous to the previous proof, we have that
\begin{eqnarray}
	\|(V^{(n)})^{-1/2}\|
	&\leq&
	K\sup_{i\in L^{(n)}}
	\left(1+\sum_{r=1}^\infty\exp(-r/\xi)|S_r(i)|\right)\\
	&\leq&
	K
	\left(1+c^{(n)}\sum_{r=1}^\infty\exp(-r/\xi)r^{d^{(n)}-1}\right)\nonumber\\
	&\leq&K(1+cLi_{1-d}(\me^{-1/\xi}))<\infty,\nonumber
\end{eqnarray}
where $Li_{1-d}$ is
the polylogarithm of degree $1-d$. As the above bound holds for all $n$, we can conclude that
\begin{equation}
	\inf\Delta E^{(n)}\ge\frac{2}{K(1+cLi_{1-d}(\me^{-1/\xi}))}>0.
\end{equation}
\qed

\subsection{Proof of Theorem 4}
In the finite-temperature case, the covariance matrix 
$\gamma(T)$
is given by
\begin{equation}
\gamma(T)=
\gamma(0)
+\left(V_p(V_xV_p)^{-1/2}G\right)\oplus
\left(V_x(V_xV_p)^{-1/2}G\right).
\end{equation}
We will now proceed by demonstrating 
the exponential decay of the
entries of the matrices $G$ and $(V_xV_p)^{-1/2}=
(V_x^{1/2} V_p V_x^{1/2})^{-1/2}$ ($V_x$ and $V_p$ are assumed
to be commuting) and will then apply the
assumption on the lattice to finally arrive at Theorem 2.
To show the exponential decay, we will now proof a generalization of
a theorem of Ref.\ \cite{Benzi} to matrices reflecting
general lattices. This latter 
work is concerned with the exponential
decay of entries of matrix functions of banded matrices.
The proof will be very similar to the one in Ref.\ 
\cite{Benzi}, only that the notion of a \text{dist}ance is different:
the \text{dist}ance from the main diagonal of a matrix versus
$\text{dist}(\cdot,\cdot)$ in the graph theoretical sense. So quite
surprisingly, the ideas of Ref.\ \cite{Benzi} carry over to
this more general case with little modifications.
Note that a mere 
naive embedding of the 
potential matrices into banded matrices 
would be insufficient to find the 
above conclusion, as for every ordering of physical systems, 
the range of the banded matrices could not be kept finite.
We nevertheless state the full proof 
here for completeness.

We denote by 
\begin{equation}
	a:=\lambda_{\text{min}}(V),\,\,\,\,\,\,
	b:=\|V\|
\end{equation}	 
the minimal and maximal eigenvalue
of a real symmetric matrix $V$ of finite range.
Let a function $f:\cc\rightarrow\cc$ be such that $f\circ\psi$ is
analytic in the interior of an ellipse\footnote{We denote its half axes by $\alpha$ and $\beta$, $\alpha>1$, $\beta>0$, $\alpha>\beta$. It is then
completely specified by the parameter $\chi=\alpha+\beta$. Note
that, if $\alpha$ is known so is $\beta$ as 
$1=\alpha^2-\beta^2$.} 
$\varepsilon_\chi$, $\chi>1$, with focii in $-1$ and $1$
 and continuous on $\varepsilon_\chi$. Furthermore suppose $(f\circ \psi)(z)\in \rr$ for $z \in \rr$. Here 
 \begin{equation}
 \psi : \cc\rightarrow\cc,\;\;\; \psi(z) =\frac{(b-a)z+a+b}{2}.
\end{equation}
We can now state the generalization of Ref.\ \cite{Benzi}.

\begin{theorem}[Exponential decay of entries of matrix functions]
Let $V=(V_{i,j})$ be a positive real symmetric matrix of finite range 
$m$,  $V_{i,j}=0$ for $\text{dist}(i,j)>m/2$, let $f:\cc\rightarrow\cc$ be such that it fulfills the above assumptions. Then there exist constants $K$ and $q$, $0\le K$, $0\le q < 1$
such that
\begin{equation}
\left|[f(V)]_{i,j}\right|
\le
K q^{\text{dist}(i,j)},
\end{equation}
where
\begin{eqnarray}
	K &:=& \max\left\{
	\|f(V)\|,
	\frac{2\chi}{\chi-1}
	\max_{z\in\varepsilon_\chi}|( f\circ \psi)(z)|
	\right\},\\
	q &:=& \left(\frac{1}{\chi}\right)^{2/m}.
\end{eqnarray}
\end{theorem}
\proof For a function $g:\cc\rightarrow\cc$ analytic in the interior of the
ellipse $\varepsilon_\chi$, $\chi>1$, continuous on $\varepsilon_\chi$,
and with $g(z)\in\rr$ for real $z$, one has
\begin{equation}
\inf\left\{
\max_{-1\leq z \leq 1}|g(z)-p(z)| : p\in P_k\right\}
\leq
\frac{2}{\chi^k(\chi-1)}\max_{z\in\varepsilon_\chi}|g(z)|,
\end{equation}
where $P_k$ denotes the set of all polynomials with real coefficients and degree less than or equal to $k$. This result is due to Bernstein (see Ref.\ \cite{Bernstein} 
for a proof) and applies to the function 
$f\circ\psi$. As $V$ is assumed to be of range $m$,
we have that $p(\psi^{-1}(V))$ is of range $km$ for all polynomials $p\in P_k$.
Thus, for $\text{dist}(i,j)>km/2$ we have 
\begin{equation}
	(p(\psi^{-1}(V)))_{i,j}=0. 
\end{equation}	
	For given $i\ne j$ we
now choose $k$ such that 
\begin{equation}
	k=\lceil 2\text{dist}(i,j)/m\rceil -1, 
\end{equation}
	i.e., we have
$k<2\text{dist}(i,j)/m\leq k+1$, which yields
\begin{eqnarray}
\left|(f(V))_{i,j}\right|&=&
\left|
((f\circ\psi) (\psi^{-1}(V)))_{i,j}
-(p(\psi^{-1}(V)))_{i,j}\right|\\
&\leq&
\left\| 
(f\circ\psi)(\psi^{-1}(V))-p(\psi^{-1}(V)) 
\right\|\nonumber\\
&=&
\max_{-1\leq z \leq 1}\left|
(f\circ\psi)(z)-p(z)\right|,\nonumber 
\end{eqnarray}
where the last equation follows from the fact that the spectrum of $\psi^{-1}(V)$ is contained in the interval $[-1,1]$. Applying Bernstein's theorem, we know
that there exists a sequence of polynomials $p_{(n)}$ of degree $k$ that satisfy 
\begin{eqnarray}
&& \lim_{n\rightarrow\infty}
\max_{-1\leq z \leq 1}
\left|
(f\circ\psi)(z)-p_{(n)}(z)\right| \nonumber\\
&& \hspace{1cm} =
\inf\left\{
\max_{-1\leq z \leq 1}|(f\circ\psi)(z)-p(z)| : p\in P_k\right\}\nonumber\\
&& \hspace{1cm} \leq
\frac{2}{\chi^k(\chi-1)}\max_{z\in\varepsilon_\chi}|(f\circ\psi)(z)|
\nonumber\\
&& \hspace{1cm} \leq 
\frac{2\chi}{\chi-1}\max_{z\in\varepsilon_\chi}|(f\circ\psi)(z)|
\left(\frac{1}{\chi}\right)^{2\text{dist}(i,j)/m}.
\end{eqnarray}
where the last inequality follows from the choice of $k$ and $\chi>1$. Along
with the fact that $(f(V))_{i,i}\leq \|f(V)\|$ this concludes the proof.
\qed

We will now apply Theorem 3 to the function 
$f$ defined as
\begin{equation}
	f(z)=\frac{2}{\me^{2\sqrt{z}/T}-1},
\end{equation}
-- reflecting the matrix function that is needed in order to
evaluate the second moments of a Gibbs state --
and the matrix $V_xV_p$ in order to 
prove Theorem 2. 
In the notation of Theorem
3 we have $V=V_xV_p$, which is of range $2m$, $a=(\Delta E/2)^2$, and $b=\| V_xV_p\|$. The function $f(z)$ 
is analytic for all 
\begin{equation}
	z\in\cc\backslash \{z\in\rr : z\leq 0\} 
\end{equation}
and therefore $f\circ\psi$ is analytic for all $z\in\cc$ with $\Re (z)<(a+b)/(b-a)$.
So for $\chi=b/(b-a)=\alpha+\beta$ the function $f$ fulfills the requirements of Theorem 3 and $|(f\circ\psi)|$ 
attains its maximum at
$z=-\alpha$. Thus,
\begin{eqnarray}
	\frac{2\chi}{\chi-1}\max_{z\in\varepsilon_\chi}|(f\circ\psi)(z)|
	&\geq&
	\frac{2}{\me^{2({a(1-a/(4b))})^{1/2}/T}-1}\\
	&\geq&
	 \frac{2}{\me^{2\sqrt{a}/T}-1}
	=\| f(V) \|.\nonumber
\end{eqnarray}
To summarize, we have
\begin{equation}
	|G_{i,j}|\leq
	\frac{4\|V_xV_p\|/\left(\Delta E/2\right)^2
	}{
	\exp\left(\frac{\Delta E}{T}
	   \left({1-\frac{\left(\Delta E/2\right)^2}
	{4\|V_xV_p\|}}\right)^{1/2}
		\right)-1}
	\left(1-\frac{\left(\Delta E/2\right)^2}
	{\|V_xV_p\|}\right)^{\text{dist}(i,j)/m}.
\end{equation}
To bound the entries of $(V_xV_p)^{-1/2}$
we could also apply Theorem 3, however, for this special
function, the bound can be given in a more straight-forward
manner. To this end, consider again the power series expansion of the square root, Eq.\ (\ref{squareroot}).
Denote by $a$ and $b$ the minimal and maximal eigenvalue
of $V=V_xV_p$, respectively. 
As matrices $V_x$ and $V_p$ are
assumed to be of range $m$, $V$ is of range $2m$. Thus, for given $i\ne j$ choosing
$k=\lceil \text{dist}(i,j)/m\rceil-1$, i.e., $k<\text{dist}(i,j)/m\leq k+1$, yields
\begin{eqnarray}
\left|\left(V^{-1/2}\right)_{i,j}\right|&=&\frac{1}{\sqrt{b}}\left|\left((V/b)^{-1/2}\right)_{i,j}-\left(\id+\sum_{r=1}^k a_r (\id-V/b)^k\right)_{i,j}\right|\nonumber\\
&\leq& \frac{1}{\sqrt{b}}
\left\| (V/b)^{-1/2}-\left(\id+\sum_{r=1}^k a_r(\id-V/b)^k\right)\right\|\nonumber \\
&=&
\frac{1}{\sqrt{b}}\max_{a/b\leq x \leq 1}\left|\sum_{r=k+1}^\infty a_r (1-x)^k\right|
\leq
\frac{1}{\sqrt{b}}\sum_{r=k+1}^\infty a_r (1-a/b)^k\nonumber\\
&=&\frac{\sqrt{b}}{a}\left(1-\frac{a}{b}\right)^{k+1}
\leq
\frac{\sqrt{b}}{a}\left(1-\frac{a}{b}\right)^{\text{dist}(i,j)/m}.
\end{eqnarray}
For diagonal terms, we have $|(V^{-1/2})_{i,i}|\leq \|V^{-1/2}\|=1/(\Delta E/2)$, i.e.,
for all $i,j$ we have
\begin{equation}
\label{decay_inv_sqr_v}
\left|\left((V_xV_p)^{-1/2}\right)_{i,j}\right|
\leq
\frac{\sqrt{\|V_xV_p\|}}{(\Delta E/2)^2}\left(1-\frac{(\Delta E/2)^2}{\|V_xV_p\|}\right)^{\text{dist}(i,j)/m},
\end{equation}
and thus, together with the assumption on the lattice,
\begin{eqnarray}
	\left|((V_xV_p)^{-1/2}G)_{i,j}\right|  \leq 
	\frac{4l_0\|V_xV_p\|^{3/2}/\left(\Delta E/2\right)^4
	}{
	\exp\left(\Delta E
	\left({1-\frac{\left(\Delta E/2\right)^2}{4\|V_xV_p\|}}\right)^{1/2}
	/T\right)-1}
	\me^{-\text{dist}(i,j)\nu}.
\end{eqnarray}
The last step is now analogous to the proof in the zero temperature
case, Eq.\ (\ref{bandedproduct}). We finally arrive at
\begin{eqnarray}
\left|\left(V_p(V_xV_p)^{-1/2}G\right)_{i,j}\right|&\leq&
\frac{4l_0\|V_xV_p\|^{3/2}v_{d,m/2}\|V_p\|/\left(\Delta E/2\right)^4
}{
\exp\left(\Delta E
\left({1-\frac{\left(\Delta E/2\right)^2}{4\|V_xV_p\|}}\right)^{1/2}
/T\right)-1}
\me^{-\nu \text{dist}(i,j)/2},\;\;\;\;\;\;\\
\left|\left(V_x(V_xV_p)^{-1/2}G\right)_{i,j}\right|&\leq&
\frac{4l_0\|V_xV_p\|^{3/2}v_{d,m/2}\|V_x\|/\left(\Delta E/2\right)^4
}{
\exp\left(\Delta E
\left({1-\frac{\left(\Delta E/2\right)^2}{4\|V_xV_p\|}}\right)^{1/2}
/T\right)-1}
\me^{-\nu \text{dist}(i,j)/2}\;\;\;\;\;\;
\end{eqnarray}
for $\text{dist}(i,j)\geq m$, which concludes the proof.\qed

\subsection{Proof of Theorem 5}
From entanglement theory we know that an upper bound to the entropy of
entanglement is given by the logarithmic 
negativity \cite{Neg,Neg2,Neg3,Neg4,Neg5,Neg6}, 
which in turn can be bounded from above by an expression
depending only on the two-point position correlation function with 
respect to the ground
state \cite{Area2},
\begin{equation}
E_S^I\leq \frac{4\| V\|^{1/2}}{\log(2)}\sum_{\substack{i\in I\\ j\in L\backslash I}}\left|\langle \hat{x}_i\hat{x}_j\rangle\right|.
\end{equation}
From the proof of  Theorem 4, Eq.~(\ref{decay_inv_sqr_v}), we know that
\begin{equation}
\left|\langle \hat{x}_i\hat{x}_j\rangle\right|=\left|\left(V^{-1/2}\right)\right|
\leq \frac{\| V\|^{1/2}}{(\Delta E/2)^2}\exp[-\text{dist}(i,j)/\xi],
\end{equation}
where
\begin{equation}
\xi=\frac{m}{\log\left(\frac{\| V\|}{\| V\|-(\Delta E/2)^2}\right)}.
\end{equation}
Thus, we find
\begin{equation}
\log(2)
E_S^I\leq \frac{4\| V\|}{(\Delta E/2)^2}\sum_{\substack{i\in I\\ j\in L\backslash I}}\exp[-\text{dist}(i,j)/\xi]=\frac{4\| V\|}{(\Delta E/2)^2}\sum_{r=1}^\infty\me^{-r/\xi}N_r,
\end{equation}
where we defined
\begin{equation}
N_r=\sum_{j\in L\backslash I}\sum_{\substack{i\in I\\ \text{dist}(i,j)=r}}\!\!\!\!\! 1.
\end{equation}
Note that, $N_1$ coincides with the definition of the surface 
area of $I$, $N_1=s(I)$.
Let us now define the "outer boundary" of $I$:
\begin{equation}
\partial I:= \left\{ j\in L\backslash I: \text{ there exists a } i\in I\text{ such that }\text{dist}(i,j)=1\right\}.
\end{equation}
We immediately find $|\partial I|\leq s(I)$ and we can restrict the sum over $L\backslash I$ to the set
\begin{equation}
A_r:=\bigcup_{i\in\partial I}\left\{ j\in L\backslash I:\text{dist}(i,j)\leq r-1\right\},\;\;
\left|A_r\right |\leq s(I)v_{d,r},
\end{equation}
i.e.,
\begin{eqnarray}
N_r&=&\sum_{j\in A_r}\sum_{\substack{i\in I\\ \text{dist}(i,j)=r}}\!\!\!\!\! 1
\leq\sum_{j\in A_r}\sum_{\substack{i\in L\\ \text{dist}(i,j)=r}}\!\!\!\!\! 1
=\sum_{j\in A_r} \left|S_r(j)\right|\nonumber\\
&\leq&cr^{d-1}v_{d,r}s(I)\leq c^2r^{2d-1}s(I).
\end{eqnarray}
To summarize,
\begin{eqnarray}
E_S^I&\leq&\frac{4\| V\|c^2}{\log(2)(\Delta E/2)^2}s(I)\sum_{r=1}^\infty\me^{-r/\xi}r^{2d-1}\nonumber\\
&=&\frac{4\| V\|c^2Li_{1-2d}(\me^{-1/\xi})}{\log(2)(\Delta E/2)^2}s(I),
\end{eqnarray}
where $Li_{1-2d}$ is the polylogarithm of degree $1-2d$.\qed
\section{Discussion and examples}

In this section, we discuss a few special cases to 
exemplify the previous results. It is important to note
that the previous results hold true in case of general
lattices, beyond structures with a very high translational 
symmetry.


\begin{example}[Disordered one-dimensional system] 
This example is to highlight that besides locality, no
assumptions on the coupling are required. In particular,
we allow for random coupling, as for example in the
following sense.
Consider a sequence of one-di\-men\-sional systems on a one-dimensional chain with
periodic boundary conditions, such that
$G^{(n)}=(L^{(n)},E^{(n)})$ is 
characterized by $L^{(n)}=[1,...,n ]$ 
and the adjacency matrix with entries
\begin{equation}
	E_{i,j}^{(n)}= \delta_{i,j+1}+ \delta_{i,j-1}+\delta_{i,n} \delta_{j,1}+
	\delta_{j,n} \delta_{i,1},\,\,\,\,\, i,j=1,...,n.
\end{equation}
The coupling is specified by $C^{(n)}=(G^{(n)},V^{(n)},\id)$.
Now let $(r_1,...,r_n)$ be a vector of realizations of 
random numbers taken from the interval $[0,1]$. Now let
$(V^{(n)})_{i,i}= 3$,	
and 
\begin{eqnarray}
	(V^{(n)})_{i,j}= \left(\delta_{i,j+1}+ \delta_{i,j-1}\right)
	r_i+\left(\delta_{i,n} \delta_{j,1}+
	\delta_{j,n} \delta_{i,1}\right) r_n,\,\,\,\,\,\,\,\, i,j=1,...,n.
\end{eqnarray}
From Gershgorin's theorem 
we then know that 
$\Delta E=2\lambda_{\text{min}}^{1/2}(V_x^{(n)})\ge 2$.
Also, 
$\|V^{(n)} \|\leq 5$ for all $n$.
Hence, Theorem 1 can be applied, and we find 
exponentially decaying correlation functions, even in this 
disordered system. Note that this example can also be 
generalized to $d$-dimensional general lattices, as
one can straightforwardly find upper and lower bounds
to the spectral values of $V$ using the same arguments 
as above. 

\end{example}

\begin{example}[Thermal states of rotating wave Hamiltonians]
\label{thermalexample}
Consider the coupling $C^{(n)}=(G^{(n)},V^{(n)},V^{(n)})$ on a general graph $G^{(n)}=(L^{(n)},E^{(n)})$. Such Hamiltonians correspond to the case
of ``rotating-wave Hamiltonians''.
For each $n$, the ground state is
easy to find: it is the product state of uncoupled degrees of 
freedom. The
covariance matrix for the zero temperature case is then given by 
\begin{equation}
	\gamma=\id\oplus\id 
\end{equation}
for all $n$. Indeed, 
the correlation length is zero.  For finite
temperature however, we find 
\begin{equation}
	\gamma(T)=\left(\id+2\left(
	\exp(2V^{(n)}/T)-\id\right)^{-1}\right)\oplus\left(\id+2\left(\exp(2V^	{(n)}/T)-\id\right)^{-1}\right).
\end{equation}
Now, e.g., choose
\begin{equation}
	V^{(n)} = \id - c E^{(n)},
\end{equation}
where $0<c<1/2$ and $E^{(n)}$ as in the previous example.
Due to the circulant structure of $V^{(n)}$, we have $\Delta E=2\sqrt{1-2c}$, $\|V\|=1+2c$, and
\begin{equation}
\langle\hat{x}_i\hat{x}_j\rangle=\langle\hat{p}_i\hat{p}_j\rangle=
\delta_{i,j}+\frac{2}{n}\sum_{k=1}^n\frac{\me^{2\pi k(i-j)/n}}{\me^{2\lambda_k(V)/T}-1}.
\end{equation}
Subsequently, we merely
sketch the argument leading to
the statement that the correlation length is non-zero
for $T>0$, but it should be clear how this can be made
a rigorous statement. 
For high temperature 
we can approximate $\exp(2\lambda_k(V)/T)-1\approx 2\lambda_k(V)/T$, and thus
\begin{equation}
\langle\hat{x}_i\hat{x}_j\rangle=\langle\hat{p}_i\hat{p}_j\rangle\approx
\delta_{i,j}+T((V^{(n)})^{-1})_{i,j}.
\end{equation}
The explicit inverse of $V^{(n)}$ is known \cite{Dow} 
and given by
\begin{eqnarray}
	((V^{(n)})^{-1})_{i,j}&=&\frac{1+q^2}{(q^n-1)(q^2-1)}\left(q^{n-|i-j|}
	+q^{|i-j|}\right),\\
 	q&:=&\frac{1-\sqrt{1-4c^2}}{2c}.
\end{eqnarray}
So, in the limit $n\rightarrow\infty$, $n$ odd, we get for $i\ne j$
\begin{equation}
\langle\hat{x}_i\hat{x}_j\rangle=\langle\hat{p}_i\hat{p}_j\rangle=
\frac{1+q^2}{1-q^2}T\exp(-\text{dist}(i,j)/\xi)\;\;\; \text{for}\;\;\; T\gg 1,
\end{equation}
where
\begin{equation}
\xi=\frac{1}{\log(1/q)}>0.
\end{equation}
Hence, the thermal states have longer-ranged
correlations as compared to the ground state. This can be explained as the 
first excited state is already no longer a product state, unlike the ground
state itself.
\end{example}

\begin{example}[Energy gap from exponential decay]
Here we will exemplify the argument leading to
a statement on the existence of a spectral gap 
for a simple one-dimensional system. 
To this end consider again the 
one dimensional chain $G^{(n)}=(L^{(n)},E^{(n)})$
with $L^{(n)}$ and $E^{(n)}$ as above and $C^{(n)}=(G^{(n)},V^{(n)},\id)$, i.e.,
$\Delta E^{(n)}=2\lambda_{\text{min}}^{1/2}(V^{(n)})$. Now
suppose that in the large system limit, $n\rightarrow\infty$, $n$ odd, either
\begin{equation}
\label{xexp}
\langle\hat{x}_i\hat{x}_j\rangle=K\exp(-\text{dist}(i,j)/\xi)
\end{equation}
or
\begin{equation}
\label{pexp}
\langle\hat{p}_i\hat{p}_j\rangle=K\exp(-\text{dist}(i,j)/\xi),
\end{equation}
with constants $K>0$ and $\xi>0$.
We can then show that the system is gapped in the following way:
Consider the matrix $A^{-1}$ (cp. the previous example) 
the entries of which are given by 
\begin{equation}
(A^{-1})_{i,j}=\frac{1+q^2}{(q^n-1)(q^2-1)}\left(q^{n-|i-j|}+q^{|i-j|}\right).
\end{equation}
That is, if  (\ref{xexp}) holds, 
we have in the limit of large $n$
\begin{equation}
\langle\hat{x}_i\hat{x}_j\rangle=K\frac{1-q^2}{1+q^2}(A^{-1})_{i,j},
\end{equation}
with $0<q=\exp(-1/\xi)<1$. Under (\ref{pexp}) the same holds for $\langle\hat{p}_i\hat{p}_j\rangle$. As we know that the inverse of $A$ is a circulant matrix, 
$A=\id-cE^{(n)}$, $c=q/(1+q^2)$, we know the minimal and maximal eigenvalue of $A$ are given 
by $\|A^{-1}\|=1+2c$, $\lambda_{\text{min}}(A)=1-2c$ for all $n$. And thus, under (\ref{xexp}) 
\begin{eqnarray}
	\Delta E&=&2\lambda_{\text{min}}(V^{1/2})
	=
		\frac{2}{\lambda_{\text{max}}(V^{-1/2})}\nonumber\\
		&=&\frac{2(1+q^2)}{K(1-q^2)\lambda_{\text{max}}(A^{-1})}
		\nonumber\\
		&=&\frac{2(1-q)^2}{K(1-q^2)}.
\end{eqnarray}
Similarly, in case that  (\ref{pexp})  
holds true, we arrive at
\begin{eqnarray}
	\Delta E &=& 2\lambda_{\text{min}}(V^{1/2})=
	2K\frac{1-q^2}{1+q^2}\lambda_{\text{min}}(A^{-1})\nonumber\\
	&=& \frac{2K(1-q^2)}{(1+q)^2}.
\end{eqnarray}
We can hence conclude in both cases that 
we have that the system is gapped,
\begin{equation}
	\Delta E\ge \frac{2(1-\exp(-1/\xi))^2}{1-\exp(-2/\xi)}
	\min\left\{K,1/K	\right\}>0.
\end{equation}
\end{example}

\begin{example}[Algebraically decaying correlation functions]
Consider a general lattice $G^{(n)}$ of dimension $d$ 
equipped
with the coupling $C^{(n)}=(G^{(n)},V^{(n)},\id)$, where we define the
translationally invariant coupling $V^{(n)}$ via, $\eta>d$, 
\begin{eqnarray}
(V^{(n)})^{-1/2}=W^{(n)},\;\;\;
(W^{(n)})_{i,j}=\left\{
\begin{array}{ll}
	\text{dist}(i,j)^{-\eta}&\text{ for } i\ne j,\\
	1+\sum_{\substack{j\in L\\ j\ne i}}
	\text{dist}(i,j)^{-\eta}
	&\text{ for } i= j.
\end{array}
\right.
\end{eqnarray}
Again, 
from Gershgorin's theorem, we know that $\lambda_{\text{min}}(W^{(n)})\ge 1$ and, together with the definition 
of the lattice dimension, Eq.\ (\ref{dimension}),
\begin{eqnarray}
	\|W^{(n)}\| &\leq & 1+2\sum_{\substack{j\in L\\ j\ne i}}
	\frac{1}{\text{dist}(i,j)^\eta}\nonumber\\
	   &\leq &
	   1 + 2 \sum_{r=1}^\infty
	   r^{-\eta} |
	   S_r(i)
	   |
	   \nonumber\\
	   &\leq &
	   1 + 2 c \sum_{r=1}^\infty
	   r^{d-1-\eta}\nonumber\\
	&=& 1+2 c \zeta(1-d+\eta)<\infty,
\end{eqnarray}
for all $n$.
The system is thus gapped and $\|V^{(n)}\|\leq 1$, i.e., all assumptions of Theorem 1 
are met except that $V^{(n)}$ is not of finite range. 
Hence we cannot conclude that the correlation functions
are exponentially decaying. Indeed, we find that 
the position correlation function 
for $i\ne j$ is given by
\begin{equation}
\langle\hat{x}_i\hat{x}_j\rangle=\frac{1}{\text{dist}(i,j)^\eta},
\end{equation}
so the correlations in this non-local system
are algebraically decaying, despite being gapped.
\end{example}

\section{Outlook}

In this paper, we have rigorously revisited the 
question of algebraically and exponentially decaying correlation functions
in gapped harmonic systems on general lattices. Both
for the ground states, as well as for Gibbs states, we
considered the implications of the gap and the coupling
strength to the correlation length. 
For systems only coupled in position, we showed 
that an energy gap 
can be deduced from algebraically decaying correlations. 
We also found an equivalence between the existence of a 
gap and exponentially decaying correlation functions
for local Hamiltonians. For zero temperature,
no assumptions have been made on
the underlying lattice. This harmonic quasi-free
case is a particularly transparent instance of a physical
system where this physically plausible connection can be
explored, explicitly making use of properties of 
matrix functions of banded matrices, instead of
exploiting Lieb-Robinson type results. As such, we 
cover cases not included in the class of
systems considered in Ref.\ \cite{Hastings}.

We showed that these statements have immediate implications 
on the scaling of entanglement in ground and thermal 
states of quantum many-body systems: under these conditions,
the geometric entropy, so the von-Neumann entropy of the
reduction, is bounded from above by a quantity that is
linear in the boundary area of the distinguished region. 
Equivalently, this quantifies the degree of entanglement
of the region with respect to the rest of the lattice. 
This result on a connection between the surface area
and the degree of entanglement can hence also be
identified on general graphs, hence establishing an
rigorous entanglement-area relationship in general 
finite-ranged gapped harmonic lattice systems.

\section{Acknowledgements}

We would like to thank M.M.\ Wolf, N.\ Schuch, J.I.\ Cirac,
M.B.\ Plenio, J.\ Drei{\ss}ig, T.\ Osborne, J.I.\ Latorre,
R.\ Orus, M.\ Fleischhauer,
and the participants of the meeting
on Quantum Information Theory in York, UK, July 2005,
for interesting discussions on the subject of the paper. 
This work has been supported by the DFG
(SPP 1116, SPP 1078),
the EU (QUPRODIS, QAP), the EPSRC, and the
European Research Councils (EURYI).

\end{document}